# "Equal bi-Vectorized" (EbV) method to high performance on GPU

Seyyed Amirreza Hashemi[1], Mohsen Lahooti[2] and Ebrahim Shirani[3]

[1] Department of Mechanical Engineering, Isfahan University of Technology; a.hashemi@me.iut.ac.ir
[2] Department of Mechanical Engineering, Isfahan University of Technology; m.lahooti@me.iut.ac.ir
[3] Department of Mechanical Engineering, Isfahan University of Technology; e.shirani@cc.iut.ac.ir

**Abstract**
Due to importance of reducing of time solution in numerical codes, we propose an algorithm for parallel LU decomposition solver for dense and sparse matrices on GPU. This algorithm is based on first bi-vectorizing a triangular matrices of decomposed coefficient matrix and then equalizing vectors. So we improve performance of LU decomposition on equal contributed scheme on threads. This algorithm also is convenient for other parallelism method and multi devices. Several test cases show advantage of this method over other familiar method.

**Introduction**

Nowadays, one of the most significant issues in computational science is to improve the solver of linear system of equations. Most of the computational codes are end up solving linear set of equations. On the other hand, the numerical modeling of physical phenomena generally leads to sparse system of equations. One of the most important problems is to find an efficient solution method for large system of linear equations represented. Parallel processing is a practical way to overcome this problem. In recent years, Graphical Processing Unit (GPU) acts important role in implementation of high performance computing. GPUs consist of hardware and software systems to provide a high degree of potential parallelism [1].

Recently vast varieties of methods have been developed to solve sparse linear system of equations. Garland et al. [2,3] primary propose a method on sparse matrix vector (SpMV) on the basis of storage requirement and computational characteristics but this method is not quite efficient, since it multiplies the kernel of computing in its algorithm. Bruggenecate et al. [4] and Liu et al. [5] also present methods for sparse system of algebraic equations on the CPU and clusters. The major problem of their method is inability to run on a single workstation or a personal computer. Wang et al. [6] present a new method based on LU factorization, their method is compatible with GPUs but the set of equations used in their method is so complex and cumbersome that it is not suitable for many tasks. Amestoy et al. [7] generalized the factorized method and present a method for sparse and dense matrixes to solve system of linear equations. In their approach, data continually exchange between CPU and GPU that result in a very large overload time in parallelization. Generally, there is not any efficient and simple scheme to implement a high performance computing system. This motivates us to explore a method based on vectorization of matrices of linear system of equations on LU decomposition solver. Also these vectors are equalized in size to achieve a high performance. This paper is organized as follows. After the introduction a short review on LU decomposition method is given which followed by our new algorithm in parallelizing LU decomposition on GPUs named "equal bi-vectorized method". The performance of our scheme is presented in the Results and discussion section and finally the paper is summarized in the Conclusion section.

**LU decomposition**

LU decomposition is one of the direct methods of solving linear equations. Since this method doesn't need repeating iterations like Gauss-Jordan method, it is completely convenient to use as a parallelism process. At first, this solver decomposes coefficients the matrix in a system of equation $AX = B$ to

$$AX = B \Leftrightarrow (LU)X = B \Leftrightarrow \\ L(UX) = B \Leftrightarrow LY = B \quad (1)$$

According to equation (1), first the set of equations $LY = B$ is solved by forward substitution followed by backward substitution step to solve $UX = B$.

**Equal bi-vectorized**

Computations on GPUs are generally based on threads computation; thread is the smallest part of GPU that programmers put their data on it. Also it is so important that equal and contributed data put threads on. Consequently, the basic idea is to present a method that provides two above mentioned features: contribution and equality. Matrix of coefficients in a diagonal dominant shape to solve linear equations is read as (2):

$$\begin{bmatrix} 1 & A_{1,2} & A_{1,3} & \cdots & & A_{1,n} \\ A_{2,1} & 1 & A_{2,3} & \cdots & & \vdots \\ \vdots & & 1 & & & \\ & & & \ddots & & A_{n-2,n} \\ & & & & 1 & A_{n-1,n} \\ A_{n,1} & \cdots & & A_{n,n-2} & A_{n,n-1} & 1 \end{bmatrix} \quad (2)$$

Then in LU decomposition approach, the coefficient matrix is decomposed to two triangular matrices (3)

$$L = \begin{bmatrix} 1 & 0 & & & & 0 \\ A_{2,1} & 1 & 0 & \cdots & & \vdots \\ \vdots & & 1 & & & \\ & & & \ddots & & \\ & & & & 1 & 0 \\ A_{n,1} & \cdots & A_{n,n-2} & A_{n,n-1} & 1 \end{bmatrix}$$

$$U = \begin{bmatrix} 1 & A_{1,2} & A_{1,3} & \cdots & & A_{1,n} \\ 0 & 1 & A_{2,3} & \cdots & & \vdots \\ \vdots & & 1 & & & \\ & & & \ddots & & A_{n-2,n} \\ & & & & 1 & A_{n-1,n} \\ 0 & & & & 0 & 1 \end{bmatrix} \quad (3)$$

where L and U are lower and upper triangular matrices respectively. Now, these matrices are divided to n vectors in the same path. Hence, square matrix of coefficient are separately contributed to 2n vectors that n vectors by each decomposed matrices i.e. the L and U matrices are divided into $L_{(i)} = \begin{bmatrix} 1 & \cdots & A_{n,i} \end{bmatrix}^T_{k \times 1}$ and $U_{(i)} = \begin{bmatrix} 1 & \cdots & A_{i,n} \end{bmatrix}_{k \times 1}$ respectively. This approach provides a scheme that is so convenient to parallel decomposition of triangular matrices especially on different threads of GPU. The arrangements of these vectors are given below.

$$L_{(1)}L_{(2)} \cdots L_{(n-2)}L_{(n-1)} A \, U_{(1)}U_{(2)} \cdots U_{(n-2)}U_{(n-1)} = I \quad (4\text{-a})$$

$$A \, U_{(1)}U_{(2)} \cdots U_{(n-2)}U_{(n-1)} = \\ \left(L_{(n-1)}\right)^{-1}\left(L_{(n-2)}\right)^{-1} \cdots \left(L_{(2)}\right)^{-1}\left(L_{(1)}\right)^{-1} \quad (4\text{-b})$$

$$U_{(1)}U_{(2)} \cdots U_{(n-2)}U_{(n-1)}L_{(n-1)}L_{(n-2)} \cdots L_{(2)}L_{(1)} = A^{-1} \quad (4\text{-c})$$

According to the left hand side of equation 4-c, the divided vectors are as Eqs. (5).

$$L^{(r)} = \begin{bmatrix} A_{r,n-1-r} & A_{r,n-2-r} & \cdots & A_{r,n-1} & A_{r,n} \end{bmatrix}^T \quad (5\text{-a})$$

$$U^{(r)} = \begin{bmatrix} A_{n-1-r,r} & A_{n-2-r,r} & \cdots & A_{n-1,r} & A_{n,r} \end{bmatrix} \quad (5\text{-b})$$

And therefore

$$A^{(r)} = \begin{bmatrix} 1 & & & & \cdots & & 0 \\ & 1 & & & & & \vdots \\ & & \ddots & & & & \\ & & & A_{r,r} & A_{r,r+1} & \cdots & A_{r,n} \\ & & & A_{r+1,r} & A_{r+1,r+1} & & \\ & & & \vdots & & \ddots & \\ & & & & & & A_{n-1,n-1} \\ 0 & \cdots & & A_{n,r} & & & A_{n,n} \end{bmatrix} \quad (5\text{-c})$$

Where the elements of these vectors are defined as (6)

$$L^{(r)} = -\frac{L^{(r-1)}}{A_{r,r}} \quad n-1-r < r < n \quad (6\text{-a})$$

$$U^{(r)} = -\frac{U^{(r-1)}}{A_{r,r}} \quad n-1-r < r < n \quad (6\text{-b})$$

$$A^{(r)} = A^{(r-1)} - \frac{L^{(r-1)} \times U^{(r-1)}}{A_{r,r}} \quad n-1-r < r < n \quad (6\text{-c})$$

Until now, matrices are just vectorized in each triangular matrix so this division leads to construction of unequal length vectors. To get a highest efficiency, we mix the vectors to make vectors equals in size, so for equalizing vectors of first and end of L matrix and first and end of U matrix combine together and sequence this approach for other vectors. Consequently triangular matrices and elements are changed as:

$$L^{(r)} = \begin{bmatrix} A_{r,1} & \cdots & A_{r,n-1-r} & \cdots & A_{r,n-1} & A_{r,n} \end{bmatrix} \quad (7\text{-a})$$

$$U^{(r)} = \begin{bmatrix} A_{1,r} & \cdots & A_{n-1-r,r} & \cdots & A_{n-1,r} & A_{n,r} \end{bmatrix} \quad (7\text{-b})$$

$$L^{(r)} = -\frac{L^{(r-1)}}{A_{r,r}} \quad 1 < r < r-1, n-r-1 < r < n \quad (7\text{-c})$$

$$U^{(r)} = -\frac{U^{(r-1)}}{A_{r,r}} \quad 1 < r < r-1, n-r-1 < r < n \quad (7\text{-d})$$

$$A^{(r)} = A^{(r-1)} - \frac{L^{(r-1)} \times U^{(r-1)}}{A_{r,r}} \quad 1 < r < r-1, n-r-1 < r < n \quad (7\text{-e})$$

By this configuration we have an equal size for each decomposed vectors. Each triangular matrix is divided to (n-1)/2 vectors and the sum of them are (n-1) separated vectors so this algorithm is leaded to fit this measure with number of thread that involve in computation of these single vectors. The time for solution of each vector is almost the same, so after a first contribution, the decomposition of matrix is happened.

**Results and discussion**

In this section the performance gained by our method is presented for sparse and dense matrices of different sizes. Our examples are performed on system as configure as a NVIDIA GTX280 with 256 single cores, CPU Intel corei7 processor running at 3.2GHz, software applicant is on OS is windows 7 64-bit, complier of this test is visual studio 2008 and using a CUDA toolkit3.2 in C programming.

First example is devoted to sparse matrices of five different sizes and the results are shown in Table 1. The table shows the run time of solution of a matrix of $(n \times n)$ on GPU and CPU and the related speed up gained.

Table 1: Result by GPU and CPU

| Matrix size | GPU, sec | CPU, sec | Speed up |
|---|---|---|---|
| 500*500 | 0.00096 | 0.0042 | 4.37 |
| 1000*1000 | 0.00188 | 0.0143 | 7.6 |
| 2000*2000 | 0.00342 | 0.0572 | 16.7 |
| 4000*4000 | 0.0072 | 0.2056 | 28.4 |
| 8000*8000 | 0.0223 | 0.9205 | 41.4 |
| 16000*16000 | 0.2106 | 10.123 | 48.1 |

The second test example is similar to the previous one but the matrices are dense rather. Table 2 shows the related results of this example in terms of GPU and CPU run times and the gained speed up.

Table 2: Result by GPU and CPU

| Matrix size | GPU, s | CPU, s | Speed up |
|---|---|---|---|
| 500*500 | 0.0074 | 0.0156 | 2.1 |
| 1000*1000 | 0.0124 | 0.0583 | 4.7 |
| 2000*2000 | 0.003 | 0.239 | 7.9 |
| 4000*4000 | 0.0758 | 1.244 | 16.4 |
| 8000*8000 | 0.483 | 13.932 | 28.8 |
| 16000*16000 | 11.03 | 376.16 | 34.1 |

Results of both examples show a very promising speed up gained by the method. In the case of sparse matrices the best speed up is 48.1 while in the case of dense matrices of same size the speed up is 34.1. In addition as it is evident from the results the speed up is increased by increasing the size of matrices in both cases of sparse and dense matrices. The reason is quiet clear since in small matrices the overload time of parallelism is comparable to operational time but by increasing the size, the operational time grows more rapidly which results in a higher speed up.

The comparison of results for sparse and dense matrices also shows that in a case of sparse matrices the speed up is 1.4-2 times of dense matrices which is due to the fact that in dense matrices, there are much more elementary operation compared to sparse matrices.

We set result of our memory transfer to and from GPU to make clearer in our parallelizing and our consideration. It is helpful to mentioned that in all results, we use shared memory efficiently to elevate our performance. The result of data transferring for dense and sparse between host and device is too near, so we put average of them that approximately calculated.

Table 3: Result average data transferring between host and device memory

| Matrix size | To GPU,s | From GPU,s |
|---|---|---|
| 500*500 | 0.00021 | 0.0001 |
| 1000*1000 | 0.00025 | 0.00012 |
| 2000*2000 | 0.00038 | 0.00014 |
| 4000*4000 | 0.00061 | 0.00016 |
| 8000*8000 | 0.00084 | 0.00019 |
| 16000*16000 | 0.0012 | 0.00025 |

Finally it is noteworthy to mention that our results obviously show the high performance of this method compared to the other methods for dense and sparse set of equations on GPU. Cuda libraries in some cases are used as a solver; the highest achievement of using this library is limited to only around 15 times over [8].

**Conclusions**

Our results show that the use of GPU for solving sparse linear equations with equal bi-vectorized method achieve up to 50 times over that CPU time solutions. These results present high performance compared to the other related works. So we pose that this method is able to use another parallel device like CPU clusters.

**Acknowledgment**

Authors would like to acknowledge the CFD group of Isfahan University of Technology for providing computational platforms for all of tests.